\documentclass[12pt]{iopart}

\usepackage{iopams}  
\usepackage[dvips]{graphicx}
\usepackage{color}
\usepackage{indentfirst}
\usepackage{hyperref}

\begin{document}

\title[]
{Exploration of the memory effect on the photon-assisted tunneling via a single quantum dot: A generalized Floquet theoretical approach}

\author{Hsing-Ta Chen}
\address{Center for Quantum Science and Engineering and Department of Physics, National Taiwan University, Taipei 106, Taiwan}

\author{Tak-San Ho}
\address{Department of Chemistry, Princeton University, Princeton, New Jersey 08544}

\author{Shih-I Chu}
\address{Department of Chemistry, University of Kansas, Lawrence, KS 66045, USA}
\address{Center for Quantum Science and Engineering and Department of Physics, National Taiwan University, Taipei 106, Taiwan}

\begin{abstract}
The generalized Floquet approach is developed to study memory effect on electron transport phenomena through a periodically driven single quantum dot in an electrode-multi-level dot-electrode nanoscale quantum device.
The memory effect is treated using a multi-function Lorentzian spectral density (LSD) model that mimics the spectral density of each electrode in terms of multiple Lorentzian functions. 
For the symmetric single-function LSD model involving a single-level dot, the underlying single-particle propagator is shown to be related to a $2\times 2$ effective time-dependent Hamiltonian that includes both the periodic external field and the electrode memory effect.
By invoking  the generalized Van Vleck (GVV) nearly degenerate perturbation theory, an analytical Tien-Gordon-like expression is derived for arbitrary order multi-photon resonance d.c. tunneling current.
Numerically converged simulations and the GVV analytical results are in good agreement, revealing the origin of multi-photon coherent destruction of tunneling and accounting for the suppression of the staircase jumps of d.c. current due to the memory effect.
Specially, a novel blockade phenomenon is observed, showing distinctive oscillations in the field-induced current in the large bias voltage limit.
\end{abstract}

\maketitle

\section{Introduction}

Electron transport of a quantum system in the presence of time-dependent external fields has been studied by various approaches, often leading to observation of new phenomena and applications. 
An early experiment conducted by Dayem and Martin in 1962 \cite{PhysRevLett.8.246} studied photon-assisted tunneling (PAT) processes in superconductor-insulator-superconductor hybrid structures, in which temporally periodic fields were applied to the source and drain.
Subsequently, Tien and Gordon proposed a theoretical model of the PAT in 1963, suggesting that a time-dependent periodic external field can produce distinctive sideband structures of the electron density in the source and drain \cite{PhysRev.129.647}. 
Particularly, in the past two decades, due to the advent of nanotechnology, the effect of time-dependent fields on the electron tunneling through nanoscale devices has been extensively investigated both experimentally and theoretically. 
On the experimental side, PAT has been studied in various nanoscale systems, including GaAs/AlGaAs quantum dots \cite{PhysRevLett.73.3443,SM212} and single-donor quantum dots in semiconductor nanostructure \cite{PhysRevB.80.165331}. 
Especially, the staircase d.c. current as a function of bias voltage (the $I-V$ characteristics) in the Coulomb blockade regime \cite{EPL7927006} and the current oscillation around zero bias voltage due to external field \cite{PhysRevB.50.2019} have been observed. 
On the theoretical side, various treatments of PAT for nanoscale devices have been formulated based especially on quantum master equation approaches \cite{SST111512,PhysRevLett.78.1536} and the scattering theory in the context of the non-equilibrium Green's function (NEGF) method \cite{PhysRevB.70.155326,PhysRevB.61.5740,PhysRevB.50.5528,PhysRevB.48.8487,PR395}, including multiple photon assisted tunneling phenomena \cite{PhysRevB.56.3591,PhysRevB.58.13007} and Non-Markovian memory effect \cite{PhysRevB.80.045309}.
The PAT absorption/emission sideband structures revealed in these studies agreed with the experimental findings.

The Floquet theory has undergone extensive development and generalization in the last few decades\cite{PR390}, including the Floquet  matrix method \cite{PR138979,PhysRevA.7.2203,PRL391195}, many-mode Floquet theorem \cite{CPL96464}, and Floquet-Liouville super-matrix formalism \cite{CPL122327}. 
More recently, it has been extended to study quantum interference of periodically driven superconducting qubits \cite{PRA79032301}. 
In addition, the generalized Van Vleck (GVV) nearly degenerate perturbation theory can be used to analytically study near-resonant multi-photon processes for few level systems \cite{JCP525977,PRA31659}.
In the past decade, the Floquet approach, within the wide-band limit, has also been widely adopted in the study of periodically driven electron transport processes involving nanoscale quantum devices \cite{PR406379}. 

\begin{figure}
\begin{center}
\includegraphics[width=5.5 cm]{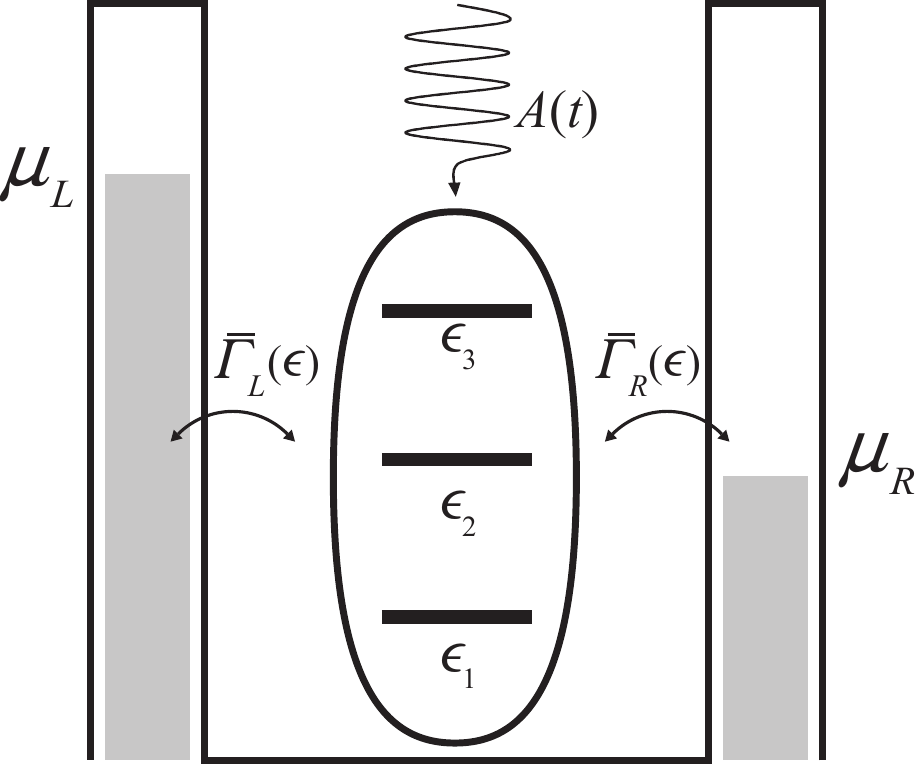}
\caption{ Energy diagram of multi-level tight-binding model of a single quantum dot in contact with two electrodes (left and right). The dot possesses discrete energy levels $\epsilon_{\mathrm{i}}$ in the presence of an external field $A(t)$. The electrode are free electron Fermi gases with the electrochemical potential $\mu_L$ and $\mu_R$ respectively. The coupling between the SQD and the electrodes are represented by the spectral density $\bar{\Gamma}_{L}(\epsilon)$ and $\bar{\Gamma}_{R}(\epsilon)$.}
\label{ED1}
\end{center}
\end{figure}

In the present work, we extend a recently formulated generalized Floquet theory, amenable to the electrode memory effect, to treat time-dependent electron transport phenomena through a periodically driven single quantum dot.
A nanoscale quantum system may contain an externally driven central junction, which usually are quantum dots \cite{PR406379,RevModPhys.75.1} or single molecules \cite{PhysRevB.70.125406,PhysRevLett.102.146801,0034-4885-64-6-201}. 
Theoretically, nanoscale quantum devices are commonly treated in terms of tight-binding models.
In the adiabatic limit, each quantum dot in the tight-binding model is usually endowed with a single electronic level that allows one electron at a time to tunnel through. 
In the general non-adiabatic situation, each quantum dot may possess multiple coupled electronic levels (as well as vibrational ones for molecules junction), thus permitting multiple electron tunneling pathways that can lead to a variety of constructive and destructive interference patterns. 
In addition, the presence of time-dependent external fields applied directly to the central junction can open up new pathways due to single- and multiple-photon resonance processes. 
The generalized Floquet approach in this paper is formulated for a single quantum dot junction endowed with $N$ noninteracting electronic states \cite{PhysRevB.50.5528} as shown in Fig.~\ref{ED1}.
This model does not permit direct electron transition between dot levels, however each level still possesses a non-zero dipole moment.
The spectral density of the electrode-junction coupling $\bar{\Gamma}_{L/R}(\epsilon)$ is represented by the Lorentzian spectral density (LSD) model in terms of a sum of multiple Lorentzian functions.
The same LSD model has also been adopted in the generalized Floquet approach developed for the electrode-multi-dots-electrode quantum system \cite{PhysRevB.79.235323}. 

The remaining parts of the paper are arranged as follows. 
In Sec.~2, we describe the general formulation and the generalized Floquet approach for the electrode-multi-level-electrode system driven by a time-periodic field. 
In Sec.~3, we consider the special case of the symmetric single-function LSD (SS-LSD) model and a single-level quantum dot. The memory effect on the multi-photon resonance PAT is studied using the GVV nearly degenerate perturbation theory.
In Sec.~4, we compare the analytical GVV results and the corresponding exact Floquet calculation, especially in the limit of large bias voltage.
Finally, a summary is given in Sec.~5.

\section{General Formulation}

\subsection{Driven multi-level model and scattering formalism}

Consider a single quantum dot (SQD) in direct contact with a left electrode (source) and a right electrode (drain), as shown in Fig.1. The Hamiltonian of the electrode-SQD-electrode system can be written as \cite{PhysRevB.70.155326,PR406379},
\begin{equation}
H(t)=H_{C}(t)+\sum_{\ell=L,R}H_{\ell}+H',
\end{equation}
where the external field is applied only on the SQD. 
Specifically, we consider a periodically driven single quantum dot that possesses $N$ non-interacting electronic states. In the tight-binding model, the Hamiltonian of the central quantum dot takes on the expression
\begin{equation}
H_{C}(t)=\sum_{\mathrm{i}=1}^{N}\epsilon_{\mathrm{i}}d^{\dagger}_{\mathrm{i}}d_{\mathrm{i}}
+A(t)\sum_{\mathrm{i}=1}^{N}\mu_{\mathrm{i}}d^{\dagger}_{\mathrm{i}}d_{\mathrm{i}},
\end{equation}
where  $\epsilon_{\mathrm{i}}$,  $d_{\mathrm{i}}$($d^{\dagger}_{\mathrm{i}}$) and $\mu_{\mathrm{i}}$ are, respectively,  the energy,  the annihilation (creation) operator and the dipole moment of the $\mathrm{i}$-th electronic state $|\mathrm{i}\rangle$ \cite{EPL7927006,Eur.Phys.J.B.68.2009}. 
The external field  $A(t)$  is a periodic function of time, i.e., $A(t+T)=A(t)$.  
The left and right electrodes are free electron Fermi gases (in thermal equilibrium) and  can be
described by the Hamiltonians
\begin{equation}
H_{\ell}=\sum_{q}\epsilon_{\ell q}c^{\dagger}_{\ell q}c_{\ell q}, \ \ell=L,R,
\end{equation}
where  $c_{\ell q}$($c^{\dagger}_{\ell q}$) is the annihilation (creation) operator for the electron states $|\ell q\rangle$ associated with the energy $\epsilon_{\ell q}$.
The contact Hamiltonian between the SQD and two electrodes is given as
\begin{equation}
H'=\sum_{\mathrm{i}=1}^{N}\left(\sum_{Lq}V_{Lq,\mathrm{i}}c^{\dagger}_{Lq}d_{\mathrm{i}}+\sum_{Rq}V_{Rq,\mathrm{i}}c^{\dagger}_{Rq}d_{\mathrm{i}}\right)+h.c.
\end{equation}
where  $V_{Lq,\mathrm{i}}$ and $V_{Rq,\mathrm{i}}$ are, respectively, the corresponding coupling parameters.

Within the framework of  the non-equilibrium Green's functions (NEGF) method,
the corresponding single-particle retarded Green's function can be computed via the relation
\begin{equation}\label{Gte}
G(t,\epsilon)=G(t+T,\epsilon)=\frac{1}{i\hbar}\int_{0}^{\infty}e^{i\epsilon t'/\hbar}U(t,t-t')dt',\ t\geq t'\geq 0,
\end{equation}
where $U(t+T,t_0+T)=U(t,t_0)$ is a $N\times N$ time-dependent matrix, representing the underlying single-particle propagator for the SQD.
The $k$th-order Fourier coefficient of the retarded Green's function $G(t,\epsilon)$ is given as
\begin{equation}\label{Gke}
G^{(k)}(\epsilon)=\frac{1}{T}\int_{0}^{T}G(t,\epsilon)e^{i k\omega t}dt,
\end{equation}
which denotes the single-particle scattering associated with the absorption or emission of $|k|$ photons at the incident energy $\epsilon$~\cite{PR406379}.
The transmission coefficient of the electron tunneling for the $|k|$-photon process is~\cite{PhysRevB.70.155326}
\begin{equation}
T_{LR/RL}^{(k)}(\epsilon)=\frac{1}{4}\sum_{\mathrm{i},\mathrm{j}=1}^{N}\bar{\Gamma}_{L/R,\mathrm{i}}(\epsilon+k\hbar\omega)\bar{\Gamma}_{R/L,\mathrm{j}}(\epsilon)\left|\langle\mathrm{i}|G^{(k)}(\epsilon)|\mathrm{j}\rangle\right|^{2},
\end{equation}
where $\bar{\Gamma}_{\ell,\mathrm{i}}(\epsilon)$ is the spectral density for the $\mathrm{i}$-th electronic state.
The total transmission coefficient of electron tunneling through the SQD is
\begin{equation}\label{Te}
T_{LR/RL}(\epsilon)=\sum_{k=-\infty}^{\infty}T_{LR/RL}^{(k)}(\epsilon)\end{equation}
The time-ensemble averaged current (i.e. the d.c. current) can be expressed in terms of the total transmission coefficients as follows
\begin{equation}\label{AVI1}
\bar{I}=\frac{e}{h}\int_{-\infty}^\infty \left\{T_{LR}(\epsilon)f_{R}(\epsilon)-T_{RL}(\epsilon)f_{L}(\epsilon)\right\}d\epsilon,
\end{equation}
where  $f_{L/R}(\epsilon)=1/\{1+e^{(\epsilon-\mu_{L/R})/k_{B}\mathcal{T}}\}$ are the corresponding Fermi-Dirac function at the temperature $\mathcal{T}$ describing the electron energy distributions in the electrodes $L$ and $R$, with $\mu_{L/R}$ being the respective electrochemical potentials. 
The bias voltage across the SQD is the difference $V=\mu_{R}-\mu_{L}$.
In the zero temperature limit, i.e. $\mathcal{T}=0$, the Fermi-Dirac function reduces to a  step function bounded by the bias voltage window $[\mu_L,\mu_R]$.

\subsection{Lorentzian spectral density model}

For a non-interacting $N$-level SQD, assuming the electrodes are initially in thermal equilibrium~\cite{PhysRevB.79.235323}, the corresponding single-particle propagator $U(t,t_0)$ is composed of $N$ independent components $U_{\mathrm{i}}(t,t_0)\equiv\langle\mathrm{i}|U(t,t_0)|\mathrm{i}\rangle$, $\mathrm{i}=1,...,N$, governed by the integro-differential equation
\begin{equation}\label{integro-diff}
i\hbar\frac{d}{dt}U_{\mathrm{i}}(t,t_{0})
=[\epsilon_{\mathrm{i}}+\mu_{\mathrm{i}}A(t)]U_{\mathrm{i}}(t,t_{0})-\frac{i}{2}\sum_{\ell=L,R}\int_{t_{0}}^{t}\bar{\Gamma}_{\ell,\mathrm{i}}(t-t')U_\mathrm{i}(t',t_{0})dt',
\end{equation}
where $\bar{\Gamma}_{\ell,\mathrm{i}}(t-t')$ is the response function (memory kernel) that can be expressed as 
\begin{equation}\label{memorykernel}
\bar{\Gamma}_{\ell,\mathrm{i}}(t-t')=\sum_{q}\frac{2}{\hbar}|V_{\ell q,\mathrm{i}}|^{2}e^{-i\epsilon_{\ell q}(t-t')/\hbar},
\end{equation}
in terms of the coupling parameters  $V_{\ell q,\mathrm{i}}$.
Here, we have taken into account the two spin states of electrons.
The spectral density $\bar{\Gamma}_{\ell,\mathrm{i}}(\epsilon)\equiv\int\bar{\Gamma}_{\ell,\mathrm{i}}(t)e^{i\epsilon t/\hbar}dt$ can be written as 
\begin{equation}
\bar{\Gamma}_{\ell,\mathrm{i}}(\epsilon)=\sum_{q}4\pi|V_{\ell q,\mathrm{i}}|^{2}\delta(\epsilon-\epsilon_{\ell q}),
\end{equation}
which a collection of delta functions at individual energy levels weighted by the corresponding coupling parameters $|V_{\ell q,\mathrm{i}}|^2$.

In the Lorentzian spectral density (LSD) model, the spectral density $\bar{\Gamma}_{\ell,\mathrm{i}}(\epsilon)$ is considered as a linear combination of $M$ Lorentzian functions, i.e.,
\begin{equation}\label{LSD}
\bar{\Gamma}_{\ell,\mathrm{i}}(\epsilon)
=\sum_{\mathrm{k}=1}^{M}\frac{a^{\ell}_{\mathrm{i}\mathrm{k}}b^{\ell}_{\mathrm{i}\mathrm{k}}}{(\epsilon-\sigma^{\ell}_{\mathrm{i}\mathrm{k}})^{2}+(b^{\ell}_{\mathrm{i}\mathrm{k}})^{2}}
=\sum_{\mathrm{k}=1}^{M}\Gamma^{\ell}_{\mathrm{i}\mathrm{k}}\frac{(b^{\ell}_{\mathrm{i}\mathrm{k}})^2}{(\epsilon-\sigma^{\ell}_{\mathrm{i}\mathrm{k}})^{2}+(b^{\ell}_{\mathrm{i}\mathrm{k}})^{2}},
\ M\geq1.
\end{equation}
where $a^{\ell}_{\mathrm{i}\mathrm{k}}$, $b^{\ell}_{\mathrm{i}\mathrm{k}}$  and $\sigma^{\ell}_{\mathrm{i}\mathrm{k}}$ are the fitting parameters for the $\mathrm{k}$-th Lorentzian function.
The spectral density $\bar{\Gamma}_{\ell,\mathrm{i}}(\epsilon)$ is usually a smooth function of energy $\epsilon$ and can be readily mimicked by a finite number of Lorentzian functions. 
Here $\Gamma^{\ell}_{\mathrm{i}\mathrm{k}}(\equiv a^\ell_{\mathrm{i}\mathrm{k}}/b^\ell_{\mathrm{i}\mathrm{k}})$, $b^\ell_{\mathrm{i}\mathrm{k}}$ and $\sigma^\ell_{\mathrm{i}\mathrm{k}}$ denote  the coupling strength,  the band-width and peak position, respectively, of the $\mathrm{k}$-th Lorentzian function for the $\mathrm{i}$-th state. 
From Eqs. (\ref{memorykernel})-(\ref{LSD}), the response function can be explicitly expressed as
\begin{eqnarray}\label{response}
\bar\Gamma_{\ell,\mathrm{i}}(t-t')
&=&\frac{1}{2\pi\hbar}\sum_{\mathrm{k}=1}^M \int \frac{a^{\ell}_{\mathrm{i}\mathrm{k}}b^{\ell}_{\mathrm{i}\mathrm{k}}}{(\epsilon-\sigma^{\ell}_{\mathrm{i}\mathrm{k}})^{2}+(b^{\ell}_{\mathrm{i}\mathrm{k}})^{2}} e^{-i\epsilon (t-t^\prime)/\hbar} d\epsilon\nonumber\\
&=&\sum_{\mathrm{k}=1}^M\frac{1}{2\hbar}a^\ell_{\mathrm{i}\mathrm{k}}  e^{-i(\sigma_{\mathrm{i}\mathrm{k}}^{\ell} -i b^{\ell}_{\mathrm{i}\mathrm{k}}) (t-t^\prime)/\hbar}.
\end{eqnarray}
In the wide-band limit (WBL), $b^{\ell}_{\mathrm{i}\mathrm{k}}\rightarrow\infty$,  thus $\bar{\Gamma}_{\ell,\mathrm{i}}(\epsilon)\rightarrow\sum_{\mathrm{k}}\Gamma^{\ell}_{\mathrm{i}\mathrm{k}}=\Gamma_{\ell,\mathrm{i}}$ is independent of energy $\epsilon$. 
As a result, we have $\bar{\Gamma}_{\ell,\mathrm{i}}(t-t')=\Gamma_{\ell,\mathrm{i}}\times\delta(t-t')$, which is free of the memory effect.

By introducing the auxiliary functions
\begin{equation}
Y^{\ell}_{\mathrm{i}\mathrm{k}}(t,t_{0})=-\frac{1}{2\hbar}\int_{t_{0}}^{t}\sqrt{a^{\ell}_{\mathrm{i}\mathrm{k}}}\exp[-i(\sigma^{\ell}_{\mathrm{i}\mathrm{k}}-i b^{\ell}_{\mathrm{i}\mathrm{k}})(t-t')/\hbar]\times U_{\mathrm{i}}(t',t_{0})dt',
\end{equation}
for $\ell=L,R$, $\mathrm{i}=1,...,N$ and $\mathrm{k}=1,...,M$, Eq.~(\ref{integro-diff}) can be recast as a set of $2M+1$ coupled ordinary differential equations of $U_{\mathrm{i}}(t,t_0)$, $Y_{\mathrm{i}}^L(t,t_0)$ and $Y_{\mathrm{i}}^R(t,t_0)$, i.e.,
\begin{equation}\label{threelevels1}
i\hbar\frac{d}{dt}
\left(
  \begin{array}{c}
    U_{\mathrm{i}}(t,t_{0}) \\
    Y^L_{\mathrm{i}}(t,t_{0}) \\
    Y^R_{\mathrm{i}}(t,t_{0}) \\
  \end{array}
\right)
=\mathcal{H}_{\mathrm{i}}(t)
\left(
  \begin{array}{c}
    U_{\mathrm{i}}(t,t_{0}) \\
    Y^L_{\mathrm{i}}(t,t_{0}) \\
    Y^R_{\mathrm{i}}(t,t_{0}) \\
  \end{array}
\right),
\end{equation}
subject to the initial conditions $U_{\mathrm{i}}(t_0,t_{0})=1$, $Y_{\mathrm{i}}^L(t_{0},t_{0})=0$, and $Y_{\mathrm{i}}^R(t_{0},t_{0})=0$. Here, the effective Hamiltonian $\mathcal{H}_{\mathrm{i}}(t)$ take on the form
\begin{equation}
\mathcal{H}_{\mathrm{i}}(t)=
\left(
  \begin{array}{ccc}
    \epsilon_{\mathrm{i}}+\mu_{\mathrm{i}}A(t) & \mathcal{Q}^L_{\mathrm{i}} & \mathcal{Q}^R_{\mathrm{i}}  \\
    \mathcal{Q}^{L\dagger}_{\mathrm{i}}  & \Sigma^L_{\mathrm{i}}-i B^L_{\mathrm{i}} & 0 \\
    \mathcal{Q}^{R\dagger}_{\mathrm{i}} & 0 & \Sigma^R_{\mathrm{i}}-i B^R_{\mathrm{i}} \\
  \end{array}
\right),
\end{equation}
where 
\begin{equation}
\mathcal{Q}^{\ell}_{\mathrm{i}}=
\left(
  \begin{array}{ccc}
    i\frac{\sqrt{a^{\ell}_{\mathrm{i}1}}}{2} & \cdots & i\frac{\sqrt{a^{\ell}_{\mathrm{i}M}}}{2}   \\
  \end{array}
\right),
\end{equation}
and 
\begin{equation}
\Sigma^{\ell}_{\mathrm{i}}=
\left(
  \begin{array}{ccc}
    \sigma^{\ell}_{\mathrm{i}1} &  & 0   \\
     & \ddots &  \\
     0 & & \sigma^{\ell}_{\mathrm{i}M} \\
  \end{array}
\right),\ 
B^{\ell}_{\mathrm{i}}=
\left(
  \begin{array}{ccc}
    b^{\ell}_{\mathrm{i}1} &  & 0   \\
     & \ddots &  \\
    0 & & b^{\ell}_{\mathrm{i}M} \\
  \end{array}
\right).
\end{equation}
The effective Hamiltonian is a time-periodic non-Hermitian matrix.
Eq.~(\ref{threelevels1}) may effectively be considered as the governing equation of a periodically driven $N$ noninteracting levels coupled to $M$ unstable levels modeling the spectral densities of states in the left and right electrodes. 
The individual level index $\mathrm{i}$ does not play a role in the following formulation and, therefore, will be dropped for simplicity. 

\subsection{Generalized Floquet approach}

By invoking the generalized Floquet theory, the solution of Eq.~(\ref{threelevels1}) can be written as
\begin{equation}
\Psi(t)=\Phi(t)\times e^{-i\Lambda t/\hbar},
\end{equation}
where $\Lambda$ is a diagonal matrix composed of complex quasienergies $\lambda_{\alpha}$ and $\Phi(t)$ is a time-dependent periodic matrix, $\Phi(t+T)=\Phi(t)$, composed of the corresponding Floquet quasi-states $\phi_{\alpha}(t)$, $\phi_{\alpha}(t+T)=\phi_{\alpha}(t)$, satisfying the eigenvalue equation,
\begin{equation}\label{QESequation}
\left(\mathcal{H}(t)-i\hbar\frac{d}{dt}\right)\phi_{\alpha}(t)=\lambda_{\alpha}\phi_{\alpha}(t).
\end{equation}
in an extended Hilbert space~\cite{PR390,PhysRevA.7.2203}.
By expanding the Floquet states $\phi_{\alpha}(t)$ 
\begin{equation}
\phi_{\alpha}(t)=\sum_{\beta=1}^{2M+1}\sum_{n=-\infty}^{\infty}\phi_{\alpha\beta}^{(n)}|\beta\rangle\times e^{i n\omega t},
\end{equation} 
Eq.~(\ref{QESequation}) can be recast as a time-independent quasienergy equation
\begin{equation}\label{qee}
\sum_{\beta=1}^{2M+1}\sum_{n=-\infty}^{\infty}\left\{\langle\alpha|\mathcal{H}^{(m-n)}|\beta\rangle-(\lambda_{\alpha}+m\hbar\omega)\delta_{\alpha\beta}\delta_{mn}\right\}\phi_{\alpha\beta}^{(n)}=0,
\end{equation}
and its concomitant adjoint equation
\begin{equation}\label{qeead}
\sum_{\beta=1}^{2M+1}\sum_{n=-\infty}^{\infty}\left\{\langle\alpha|\mathcal{H}^{\dagger (m-n)}|\beta\rangle-(\lambda^{*}_{\alpha}+m\hbar\omega)\delta_{\alpha\beta}\delta_{mn}\right\}\bar{\phi}_{\alpha\beta}^{(n)}=0,
\end{equation}
for $\alpha=1,...,2M+1$ where $|\alpha\rangle$ and $|\beta\rangle$ indicate the orthogonal unperturbed eigenstates with $A(t)=0$ and $\mathcal{H}^{(n)}=\frac{1}{T}\int_0^T\mathcal{H}(t)e^{i n\omega t}dt$. 

The fundamental solution $\mathcal{U}(t,t_0)=\Psi(t)\Psi^{-1}(t_0)=\Phi(t)e^{-i\Lambda(t-t_0)}\Phi^{-1}(t_0)$ can be obtained by solving Eq.~(\ref{QESequation}).
Given the initial condition, $U_{\mathrm{i}}(t_0,t_0)=1$, the desired single-particle propagator can be obtained $U_{\mathrm{i}}(t,t_0)=\mathcal{U}_{11}(t,t_0)$.
In practical, we can solve the quasienergy equation numerically by truncating the Floquet Hamiltonian.

\section{Symmetric single-function LSD (SS-LSD) model and a single-level quantum dot}

In this section, we consider the symmetric single-function LSD model and  a single-level quantum dot.
In particular, GVV nearly-degenerate perturbation theory is adopted to derive a Tien-Gordon-like expression for the d.c. current at single- and multi-photon resonance conditions. 
The resultant derivations can be readily extended to non-interacting multiple level cases.

\begin{figure}
\begin{center}
  \includegraphics[width=10.0 cm]{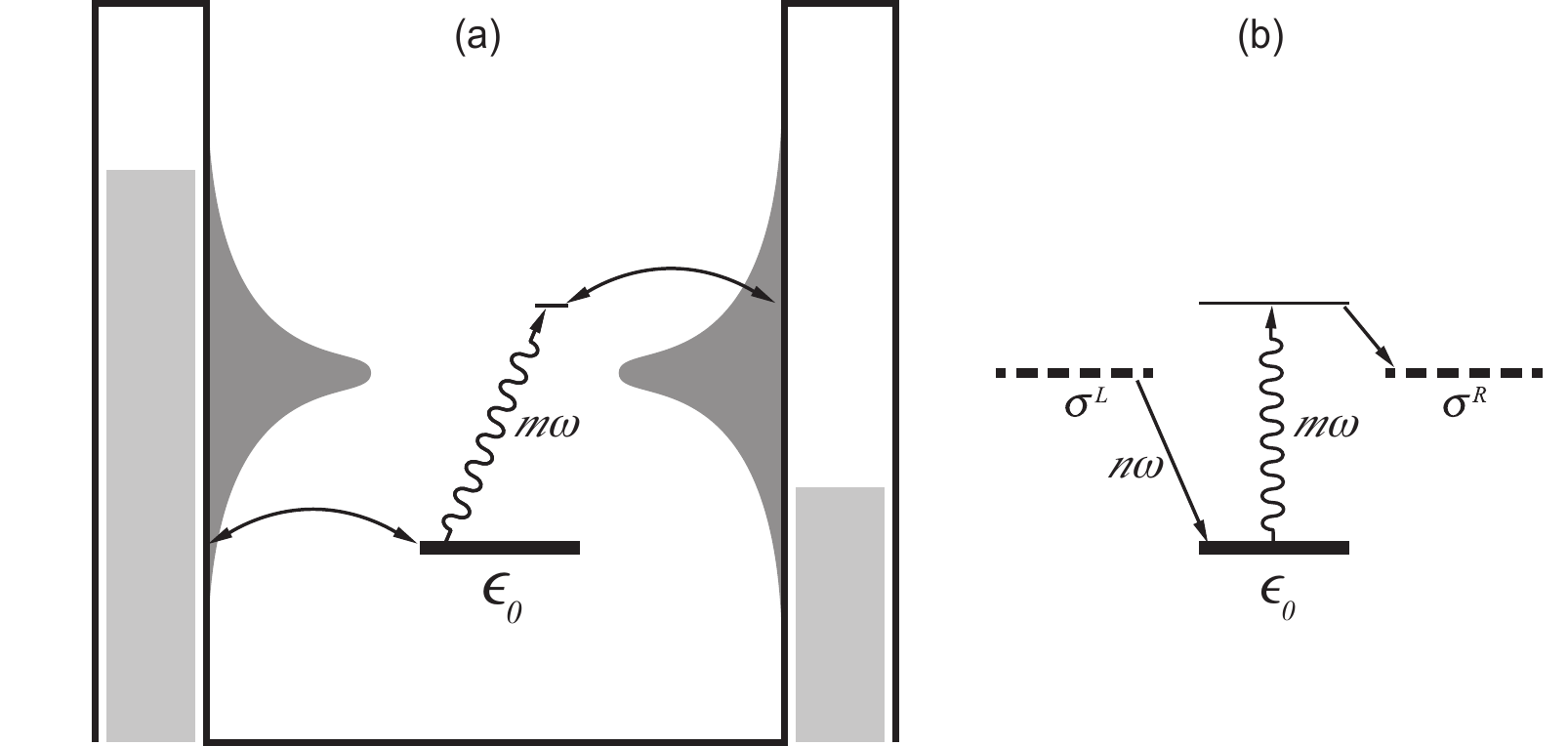}
  \caption{Energy diagram of the SS-LSD model and a driven single-level quantum dot are shown in (a), where the shade areas represent the single Lorentzian spectral density.  The effective three-level system (b) are consist of effective levels (thick dash lines), representing the center of Lorentzian function: $\sigma^L=\sigma^R=\sigma$. The  $n$-photon resonance condition and $m$-photon absorption inelastic scattering process are also shown.}\label{ED2}
\end{center}
\end{figure}

\subsection{General formulation}
We assume that the single dot level has the energy $\epsilon_0$ and dipole moment $\mu_0=1$.
The SS-LSD model depicts the left and right spectral densities with the same Lorentzian function, i.e.,
\begin{equation}\label{gamma}
\bar{\Gamma}_{L}(\epsilon)=\bar{\Gamma}_{R}(\epsilon)=\bar{\Gamma}(\epsilon)=\frac{a b}{\epsilon^{2}+b^{2}}=\Gamma\frac{b^2}{\epsilon^{2}+b^{2}},
\end{equation}
where $\Gamma=a/b$. Within the SS-LSD model, Eq.~(\ref{threelevels1}) is reduced to
\begin{equation}\label{rGSE}
i\hbar\frac{d}{dt}
\left(
\begin{array}{c}
    U(t,t_0) \\
    Y^{+}(t,t_0) \\
  \end{array}
\right)
=
\left(
  \begin{array}{cc}
    \epsilon_{0}+A(t) & i\sqrt{a/2} \\
    -i\sqrt{a/2} & \sigma-i b \\
  \end{array}
\right)
\left(
\begin{array}{c}
    U(t,t_0) \\
    Y^{+} (t,t_0)\\
  \end{array}
\right),
\end{equation}
where $Y^{\pm}(t,t_{0})=(Y^{R}(t,t_{0})\pm Y^{L}(t,t_{0}))/\sqrt{2}$ and $A(t)=A\cos\omega t$.
Noting that $Y^{-}(t,t_{0})=0$ $\forall t\geq t_0$ since it is decoupled from both $U(t,t_0)$ and $Y^+(t,t_0)$ and possesses the initial condition $Y^{-}(t_{0},t_{0})=0$. 
Consequently, the transmission coefficient and the d.c. current can be written, respectively, as
\begin{equation}\label{Te2}
T_{RL}(\epsilon)=T_{LR}(\epsilon)=T(\epsilon)=\frac{1}{4}\sum_{k}\bar{\Gamma}(\epsilon+k\hbar\omega)\bar{\Gamma}(\epsilon)|G^{(k)}(\epsilon)|^{2}
\end{equation}
and
\begin{equation}\label{AVI3}
\bar{I}=\frac{e}{h}\int_{-\infty}^\infty T(\epsilon)[f_{R}(\epsilon)-f_{L}(\epsilon)]d\epsilon.
\end{equation}
It is seen that in the WBL the transmission coefficient $T(\epsilon)$ can be further reduced to
\begin{equation}
T_{WBL}(\epsilon)=\frac{1}{4}\Gamma^2 \sum_{k}|G^{(k)}(\epsilon)|^{2},
\end{equation}
for ${\bar\Gamma}(\epsilon)\rightarrow \Gamma=a/b$ as $b\rightarrow \infty$, cf. Eq.~(\ref{gamma}).

Near the $n$-photon resonance condition $\sigma-\epsilon_0\approx n\hbar\omega$ and introducing the rotating frame
\begin{equation}\label{rotationRn}
\left(
\begin{array}{c}
    U(t,t_0) \\
    Y^+ (t,t_0)\\
  \end{array}
\right)=
\mathcal{R}(t)
\left(
\begin{array}{c}
    X'(t,t_0) \\
    Y' (t,t_0)\\
  \end{array}
\right)
\end{equation}
where
\begin{equation}
\mathcal{R}(t)\equiv
\left(
  \begin{array}{cc}
    e^{-i{\Omega}(t)} & 0 \\
    0 & e^{-i n\omega t} \\
  \end{array}
\right)
\end{equation}
with ${\Omega}(t)=\frac{1}{\hbar}\int^{t}A(t')dt'=\frac{A}{\hbar\omega}\sin\omega t$, Eq.~(\ref{rGSE}) can be transformed to 
\begin{equation}\label{rGSE'}
i\hbar\frac{\partial}{\partial t}
\left(
\begin{array}{c}
    X'(t,t_0) \\
    Y'(t,t_0)\\
  \end{array}
\right)
=\left\{\mathcal{H}_0+\xi\mathcal{V}(t)\right\}
\left(
\begin{array}{c}
    X'(t,t_0) \\
    Y' (t,t_0)\\
  \end{array}
\right)
\end{equation}
where
\begin{equation}\label{H0n}
\mathcal{H}_0=
\left(
  \begin{array}{cc}
    \epsilon_{0} & 0\\
    0 & \sigma-n\hbar\omega-i b\\
  \end{array}
\right),
\end{equation}
and 
\begin{equation}
\mathcal{V}(t)=
\left(
  \begin{array}{cc}
    0 & i e^{i(\Omega(t)-n\omega t)}\\
    -i e^{-i(\Omega(t)-n\omega t)} & 0\\
  \end{array}
\right).
\end{equation}
Here, $\xi=\sqrt{a/2}$  is a small parameter associated with the weak coupling and finite band width.
Consequently, the single-particle propagator can be computed using the relation
\begin{equation}\label{UU}
U(t,t')=\left[\mathcal{R}(t) \mathcal{U}'(t,t^\prime)\mathcal{R}^{\dagger}(t')\right]_{11}.
\end{equation}
Invoking the relation $e^{i x\sin\omega t}=\sum_{m}J_{m}(x)e^{i m\omega t}$, $x=A/\hbar\omega$ and $J_{m}(x)$ is the $m$-th order Bessel function of the first kind, $\mathcal{V}(t)$ can be Fourier expanded as $\mathcal{V}(t)=\sum_m\mathcal{V}^{(m)} e^{-i m\omega t}$,
\begin{equation}\label{Vn}
\mathcal{V}^{(m)}=
\left(
  \begin{array}{cc}
    0 & i J_{n-m}(x) \\
    -i J_{n+m}(x) & 0 \\
  \end{array}
\right).
\end{equation}
At the $n$-photon resonance condition, the off-diagonal part of the rotated Hamiltonian contain the couplings $J_{n\pm m}(x)$ corresponding to the $|m|$-photon absorption (emission), as shown in Fig.~\ref{ED2}.

\subsection{Generalized Van Vleck (GVV) nearly-degenerate perturbation theory}

Near the $n$-photon resonance condition $\sigma-\epsilon\approx n\hbar\omega$, with the aid of the generalized Van Vleck (GVV) perturbation method (see Appendix), Eq.~(\ref{rGSE'}) can be further approximated as
\begin{equation}\label{GVVn}
i\hbar\frac{\partial}{\partial t}
\left(
\begin{array}{c}
    X'(t,t_0) \\
    Y'(t,t_0)\\
  \end{array}
\right)
=\mathcal{H}_{GVV}
\left(
\begin{array}{c}
    X'(t,t_0) \\
    Y'(t,t_0)\\
  \end{array}
\right),
\end{equation}
where the $n$th-order resonance GVV Hamiltonian takes on the form 
\begin{equation}\label{GVVHn}
\mathcal{H}_{GVV}=
\left(
  \begin{array}{cc}
    \epsilon_{0}-\xi^{2}\delta_{n} & i\xi J_n(x) \\
    -i\xi J_n(x) & \epsilon_{0}-i b+\xi^{2}\delta_{n}  \\
  \end{array}
\right)
\end{equation}
in which the off-diagonal terms $\pm i\xi J_n(x)$ are the first-order correction with respect to the small parameter $\xi$ while the diagonal terms are the second order correction with 
\begin{equation}
\delta_{n}=\left\{\begin{array}{lll}
i\sum\limits_{m\neq0}\frac{b}{(m\hbar\omega)^{2}+b^{2}}J_m^2(x)&\mbox{if}&n=0\\
\sum\limits_{m\neq0}\frac{m\hbar\omega}{(m\hbar\omega)^{2}+b^{2}}J^{2}_{n+m}(x)
+i\sum\limits_{m\neq0}\frac{b}{(m\hbar\omega)^{2}+b^{2}}J^{2}_{n+m}(x)&\mbox{if}&n\neq 0\\
\end{array}\right..
\end{equation}
It can be seen that the real part of the second order correction $\xi^2\delta_n$ is responsible for the level shift \cite{PRA79032301} and the imaginary part, together with the first order correction $i\xi J_n(x)$, is responsible for the narrowing of the PAT sidebands. 

As shown in Appendix, Eq.~(\ref{Gke3}), the underlying retarded Green's function can be written as
\begin{equation}\label{Gke4}
G^{(k)}(\epsilon)\approx\sum_{m}
\frac{J_m(x)J_{m+k}(x)}{\epsilon-\epsilon_{0}-m\hbar\omega+\Delta_n+i\Xi_n},
\end{equation}
where the level shift is 
\begin{equation}\label{Delta}
\Delta_{n}\equiv\left\{\begin{array}{lll}
0&\mbox{if}& n=0\\
\frac{\Gamma}{2}\sum\limits_{m\neq 0}\frac{b m\hbar\omega}{(m\hbar\omega)^{2}+b^{2}}J^{2}_{n+m}(x)&\mbox{if}& n\geq 1\\
\end{array}\right.
\end{equation}
and the corresponding width is
\begin{equation}\label{Sigma}
\Xi_{n}=\frac{\Gamma}{2}\sum_{m=-\infty}^{\infty}\frac{b^2}{(m\hbar\omega)^{2}+b^{2}}J^{2}_{n+m}(x)\leq\frac{\Gamma}{2}=\frac{a}{2b}.
\end{equation}
Furthermore, by substituting Eq.~(\ref{Gke4}) into Eq.~(\ref{Te}), we can derive an analytical expression for the transmission coefficient 
\begin{equation}\label{TeGVV}
T_{GVV}(\epsilon)=\frac{1}{4}\sum_{m=-\infty}^{\infty}J^{2}_{m}(x)\times
\frac{\bar{\Gamma}(\epsilon)\gamma_{m}(\epsilon)}
{(\epsilon-\epsilon_{0}-m\hbar\omega+\Delta_{n})^{2}+\Xi_{n}^{2}},
\end{equation}
where the tunneling rate of the $m$-th PAT sideband is 
\begin{equation}\label{tunnelrate}
\gamma_{m}(\epsilon)
=\sum_{k=-\infty}^{\infty}\bar{\Gamma}(\epsilon+k\hbar\omega)J_{m+k}^{2}(x)
=\Gamma \sum_{k=-\infty}^{\infty}\frac{b^2J_{m+k}^{2}(x)}{(\epsilon+k\hbar\omega)^2+b^2}.
\end{equation}
It is seen that in the WBL $\gamma_m(\epsilon)\rightarrow\Gamma$, $\Delta_n\rightarrow 0$ and $\Xi_n\rightarrow\frac{\Gamma}{2}$), leading to
\begin{equation}\label{TeWBL}
T_{WBL}(\epsilon)=\frac{1}{4}\sum_{m=-\infty}^{\infty}J_m^2(x)\times
\frac{\Gamma^{2}}{(\epsilon-\epsilon_{0}-m\hbar\omega)^{2}+\frac{\Gamma^{2}}{4}}.
\end{equation}
Both $T_{WBL}(\epsilon)$ and $T_{GVV}(\epsilon)$ are composed of a collection of sidebands. 
However, the GVV sidebands exhibit additional shifts ($-\Delta_n$) and are also narrower in width ($\Xi_n\leq\frac{\Gamma}{2}$) due to the memory effect.

From Eqs.~(\ref{AVI3})~and~(\ref{TeGVV}), a Tien-Gordon-like formula for the corresponding d.c. current can be written as 
\begin{equation}\label{AVI5}
\bar{I}_{GVV}=\sum_{m=-\infty}^\infty J^{2}_{m}(x)\times\bar{I}_m,
\end{equation}
which is composed of an infinite number of weighted contributing partial currents
\begin{equation}\label{AVI6}
\bar{I}_m=\frac{e}{4h}\int_{-\infty}^{\infty} \frac{\bar{\Gamma}(\epsilon)\gamma_{m}(\epsilon)}{(\epsilon-\epsilon_{0}-m\hbar\omega+\Delta_{n})^{2}+\Xi_{n}^{2}}
\times[f_{R}(\epsilon)-f_{L}(\epsilon)]d\epsilon,
\end{equation}
for $\ m=-\infty,\cdots, +\infty$.
For the $N$-level quantum dot, the total d.c. current can be expressed as a summation of individual currents through different states.

\section{Results and discussions}

In this section, we present and discuss the memory effect on the electron transport processes of the electrode-SQD-electrode device in the presence of a periodical field, based on both the numerically converged results and approximate GVV results for the special case of symmetric single-function LSD. 
First,  we study the level shifting and width narrowing of the PAT sidebands as a function of the external field amplitude.
Second, we compare the transmission coefficient for the SS-LSD plus a single-level quantum dot and that in the wide-band limit (WBL).
Third, we study the memory effect on the staircase jumping of d.c. current.
Finally, we show the field-induced current oscillations as a function of the gate voltage (i.e. $\epsilon_0$) at the large bias voltage.
In our calculations, the frequency of the driving field ($A(t)=A\cos\omega t$) is fixed at $\hbar\omega=10~eV$, which is also used as the unit of energy, and the field amplitude $A$ is allowed to take on arbitrary values. 
The electrode-single dot coupling strength is chosen as $\Gamma^L=\Gamma^R=\Gamma=0.1~\hbar\omega$.

\begin{figure}
\begin{center}
  \includegraphics[width=12.0 cm]{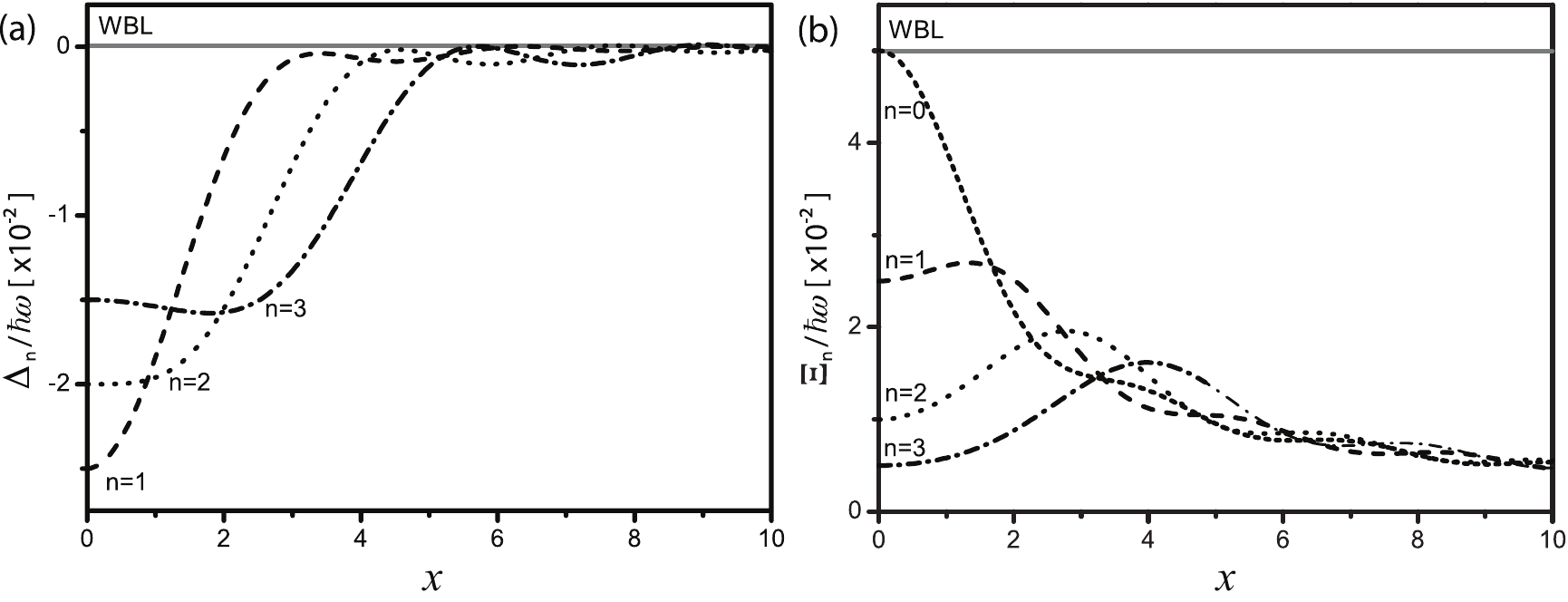}
  \caption{The amplitude dependence for (a) the level shifting $\Delta_{n}$ and (b) the sideband width $\Xi_{n}$ are plotted with respect to various resonance conditions $n=(\sigma-\epsilon_0)/\hbar\omega$ for the SS-LSD model with $\sigma=0$ and $b=1~\hbar\omega$ . The thick gray line indicate the corresponding values in the wide-band limit (i.e. $b\rightarrow\infty$).}
  \label{DNSN}
\end{center}
\end{figure}

Figure~\ref{DNSN} shows the results based on Eqs.~(\ref{Delta}) and (\ref{Sigma}), depicting the memory effect on the level shifting $\Delta_n$ and width narrowing $\Xi_n$, as a function of the amplitude $x$ ($x=A/\hbar\omega$),
at the $n=0,1,2,3$-photon resonance conditions.
It is found that in the small amplitude limit $x\ll1$, both $\Delta_n$ and $\Xi_n$ are dominated by the term $|J_0(x)|^2$, corresponding to $m=-n$ in Eqs.~(\ref{Delta}) and (\ref{Sigma}). 
At $x=0$,  the level shifting and width narrowing can be computed as
\begin{equation*}
\Delta_n=\frac{\Gamma}{2}\frac{-nb\hbar\omega}{(n\hbar\omega)^2+b^2} \leq 0\ \mbox{if}\ n\geq 1
\end{equation*}
and 
\begin{equation*}
\Xi_n=\frac{\Gamma}{2}\frac{b^2}{(n\hbar\omega)^2+b^2}\geq 0\ \mbox{for all}\ n
\end{equation*}
respectively. 
In the large amplitude limit $x\gg1$, $|J_m (x)|^2\approx (2/\pi x)\cos^2[x-(2m+1)\pi/4]$, thus $\Delta_n (n\neq 0)$ and $\Xi_n$ are inversely proportional to $x$ -- they both become smaller as $x$ becomes larger, with the former goes to zero much faster because of the cancellations between opposite signs arising from the summation index $m$ in Eq.~(\ref{Delta}).  
The observed behaviors of the level shifting and width narrowing, as shown in Fig.~\ref{DNSN} are important to understand the properties of the electron transport transmission coefficients and the resultant d.c. currents that are to be discussed below.

\begin{figure}
\begin{center}
  \includegraphics[width=16.0 cm]{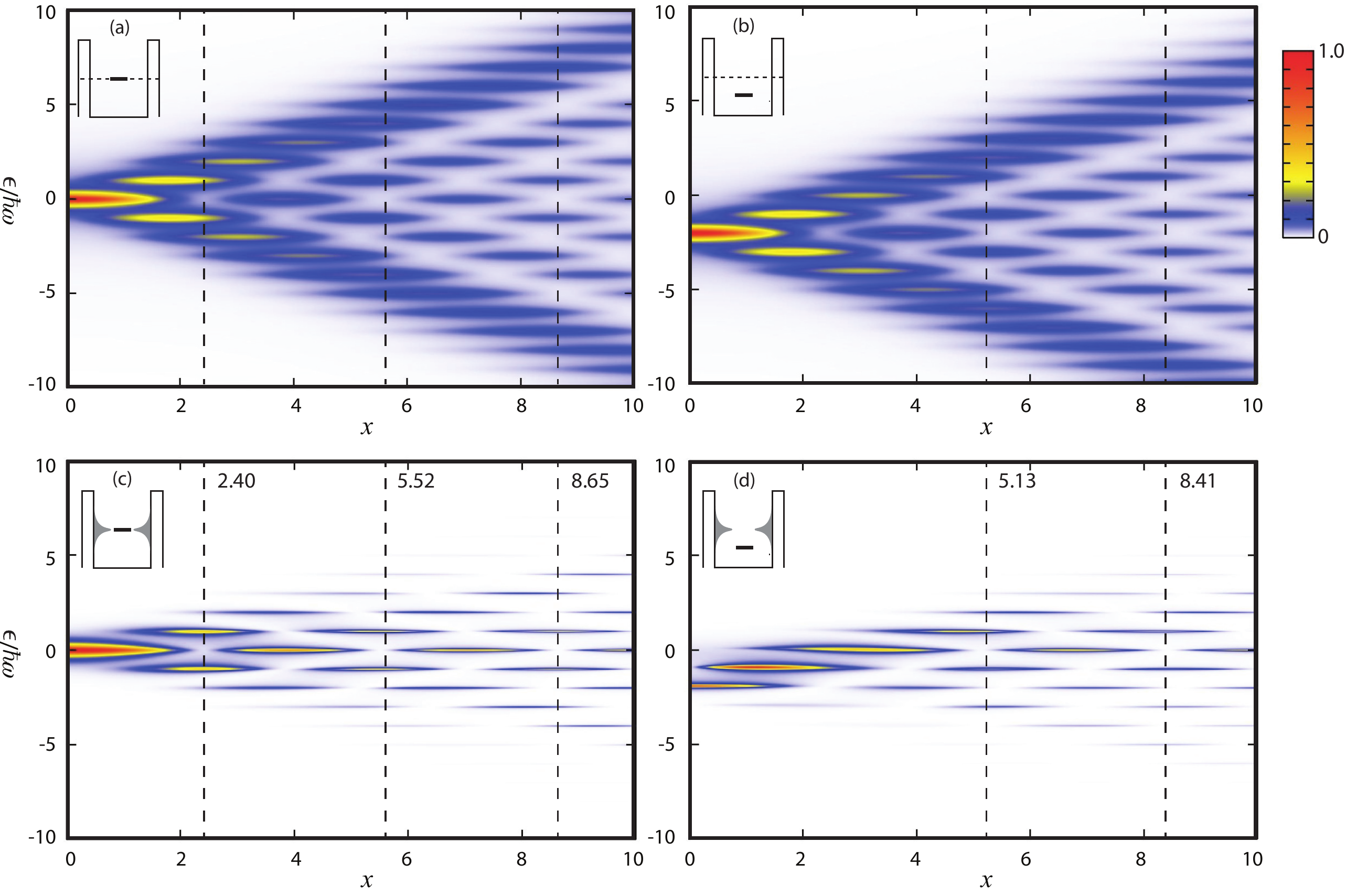}\\
  \caption{The contour plots of the transmission coefficients $T(\epsilon)$ are plotted as a function of the incident energy $\epsilon$ and the external field amplitude $x=A/\hbar\omega$ for the WBL ((a) and (b)) and the SS-LSD model with $b=2~\hbar\omega$ ((c) and (d)). We compare the WBL and SS-LSD results for the resonance conditions $n=0$ and $n=2$. The inserts show the corresponding energy diagrams. The vertical dash lines indicate the roots of $J_0(x)$ (left) and $J_2(x)$ (right) that are related to coherent destruction of tunneling.}
  \label{TXAYE}
\end{center}
\end{figure}

Depicted in Fig.~\ref{TXAYE} are contour plots of the electron transport transmission coefficients, as a function of the incident electron energy $\epsilon$ and the field amplitude $x$ for the SS-LSD parameters $\sigma=0$ (spectral peak center) and $b=2~\hbar\omega$ (spectral width).
The calculations were done based on Eqs.~(\ref{qee}) and (\ref{qeead}). 
It is found that the PAT sideband structure contains patterns of narrow peaks at the multiple integers of the applied field frequency $\hbar\omega$, i.e., $\epsilon=\epsilon_0+k\hbar\omega$, $ k=0, \pm1,\pm2,\cdots$, corresponding to various multi-photon absorption and emission processes.
Furthermore,  in the WBL (Figs. \ref{TXAYE}(a)~and~(b)), the transmission coefficient contours display the same sideband structures that are centered at different resonance energies, $\epsilon=0\ \hbar\omega$ ($n=0$) and $\epsilon=2\hbar\omega$ ($n=2$), respectively, cf. Eq.~(\ref{TeWBL}).
However, in the SS-LSD model (Figs. \ref{TXAYE}(c)~and~(d)), the transmission coefficients reveal a very different sideband patterns for different incident energies -- here the SS-LSD sidebands are symmetric for $n=0$ but asymmetric  for $n=2$, showing the results of the level shifting and width narrowing because of the memory effect, cf. Eq.~(\ref{TeGVV}).
 
\begin{figure}
\begin{center}
  \includegraphics[width=13.0 cm]{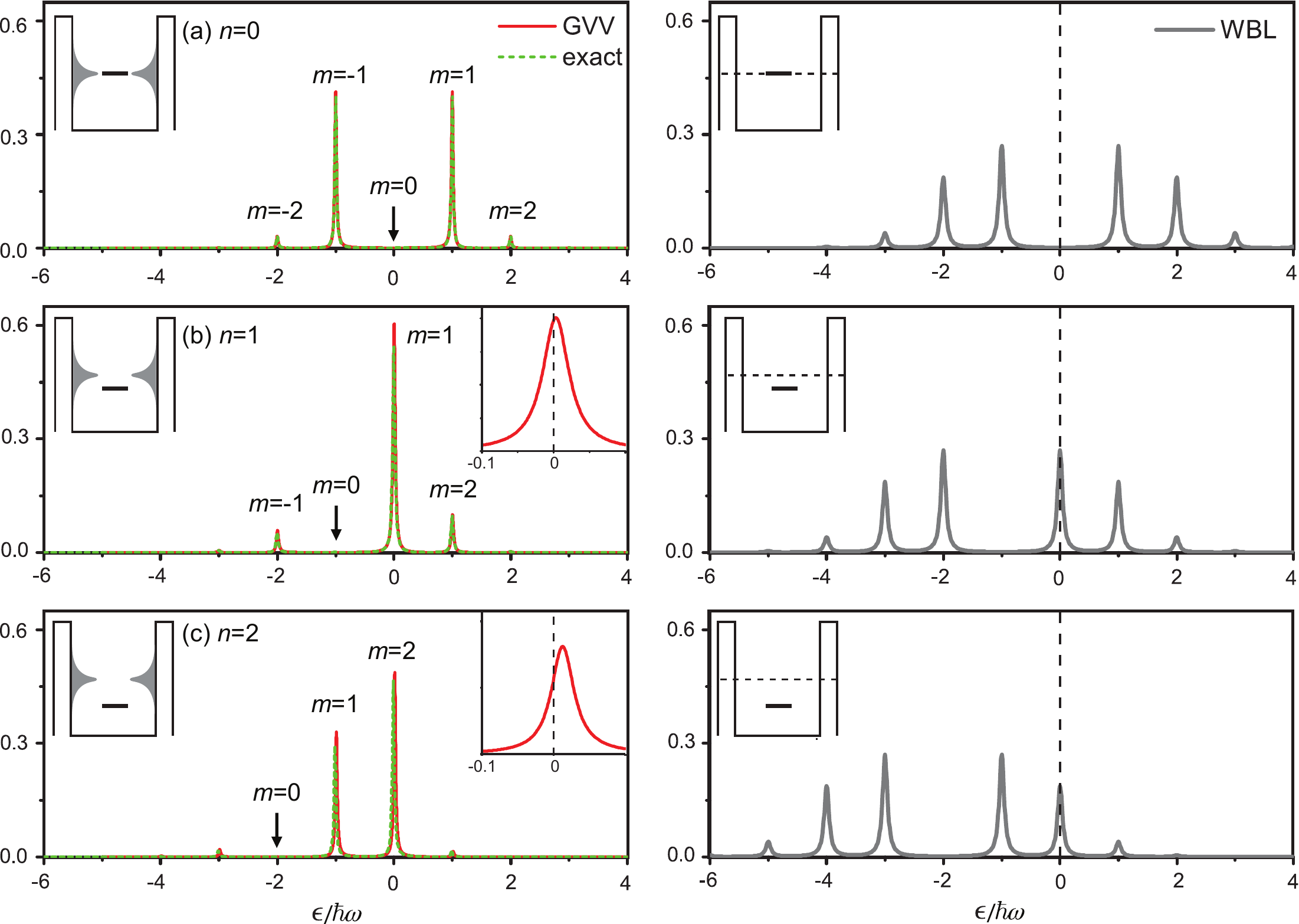}\\
  \caption{The numerical (green dashed curves) and GVV (red solid curves) results of transmission coefficients are shown as a function of the incident energy $\epsilon$ for the SS-LSD model ($b=1~\hbar\omega$) with respect to various resonance conditions  (a)~$n=0$, (b)~$n=1$, (c)~$n=2$. The corresponding energy diagrams are as the inserts and the external field amplitude is $x=2.40$. We label each peak by the $m$ index , including the vanishing peaks at $m=0$ due to CDT. The scaled insets on the right corner in (b) and (c) indicate the level shifting near $\epsilon=0$. The corresponding transmission coefficients for the WBL are shown in the right panels.}
  \label{SFETvsGVV}
 \end{center}
\end{figure}
 
Figure~\ref{SFETvsGVV} further shows numerical and GVV (green-dashed and red-solid, respectively, in the left panels) as well as the WBL calculations (the right panels) of the non-overlapping peaks ($m=0,\pm 1,\pm 2,\cdots$) in the transmission coefficients as a function of the incident energy $\epsilon$ for $n=0,1,2$.
It is found that the PAT sideband is completely suppressed at $m=0$ (i.e., $\epsilon=-n\hbar\omega$), a manifestation of the CDT phenomenon~\cite{EPL7927006,PhysRevB.79.235323}.
The inserts on the upper right corners in the panels (b) and (c) display the enlarged peaks near $\epsilon=0$, revealing significant memory effect. 
Clearly, as shown in the left panels of Fig.~\ref{SFETvsGVV}, each $n$-photon ($n=0,1,2$) resonance PAT transmission coefficient is composed of a sequence of non-overlapping sidebands ($m=0,\pm1,\pm 2,\cdots$), respectively, corresponding to $\epsilon/\hbar\omega=-n+m$ with the amplitudes approximated as
\begin{equation*}
\gamma(n,m)\times\frac{\Gamma^2}{4(\Delta_n^2+\Xi_n^2)}\times\frac{J^2_m(x)}{(-n+m)^2+1},
\end{equation*}
where
 \begin{equation*}
 \gamma(n,m)\equiv\sum_{k=-\infty}^\infty \frac{J^2_{m+k}(x)}{(-n+m+k)^2+1}.
 \end{equation*}
The $n=1,2$-photon PAT sidebands in Figs.~\ref{SFETvsGVV}(b) and~\ref{SFETvsGVV}(c) are not symmetric about $m=0$ due to the asymmetric weights of  $\gamma(n,m)$ and $\gamma(n,-m)$. 
This is in contrast to the always symmetric sideband structures about $m=0$ in the WBL, as shown in the right panel in Fig.~\ref{SFETvsGVV}.

\begin{figure}
\begin{center}
  \includegraphics[width=9.0 cm]{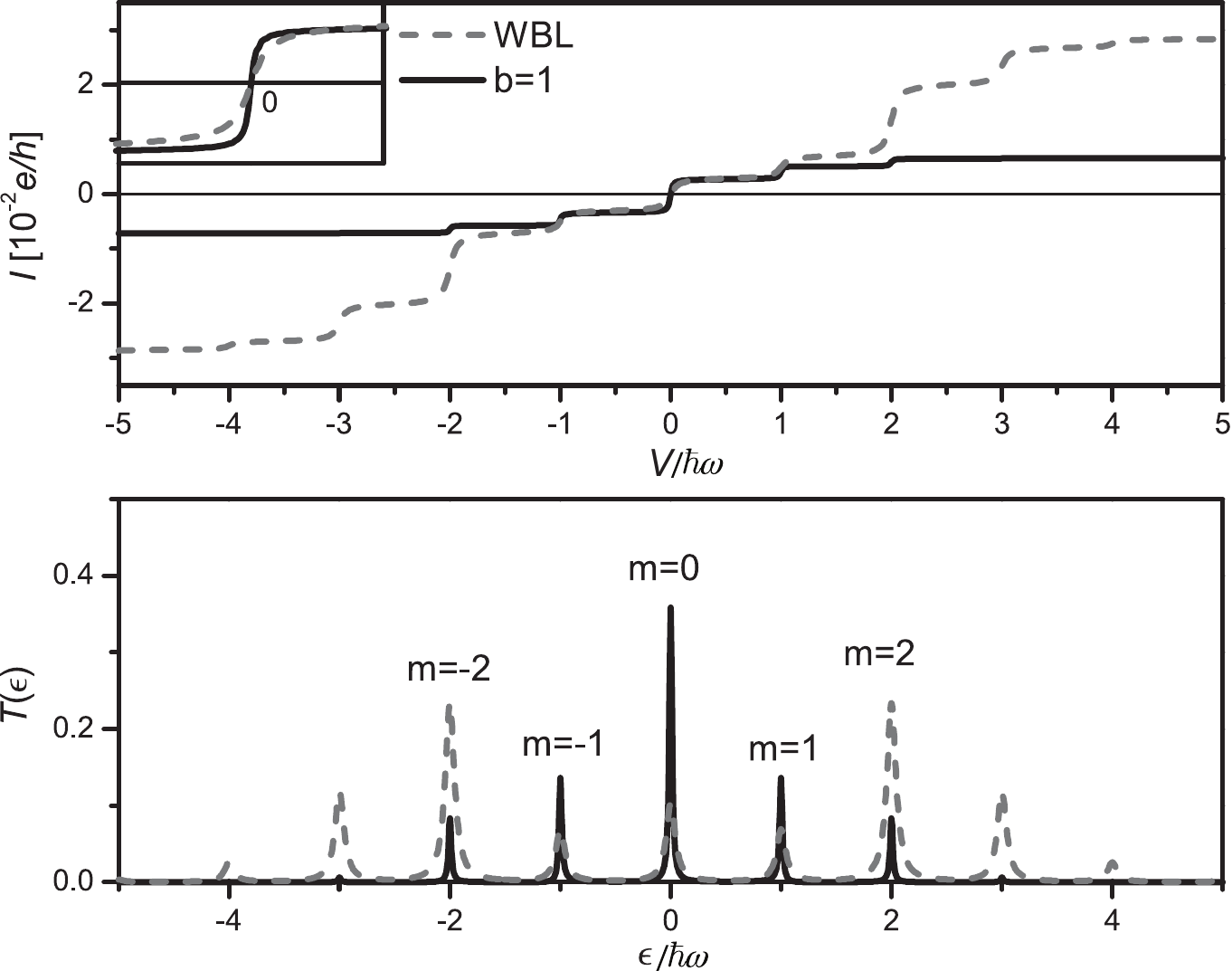}\\
  \caption{The d.c. current $\bar{I}$ is plotted as a function of bias voltage  $V=\mu_{R}-\mu_{L}$ with fixed $\mu_{L}=0$ in the upper panel and its corresponding transmission coefficient is shown in the lower panel. The results for the SS-LSD model ($\epsilon_0=\sigma=0$ and $b=1~\hbar\omega$) is compared with the WL results (gray dash curves) at $x=2~\hbar\omega$.}
  \label{IMTE}
\end{center}
\end{figure}

In Fig.~\ref{IMTE}, the upper panel shows the staircase feature of the d.c. current $\bar{I}$ as a function of the bias voltage $V=\mu_R-\mu_L$ (here $\mu_L=0$), whereas the lower panel shows the sideband structure of the underlying transmission coefficient $T(\epsilon)$ as a function of the incident energy $\epsilon$.
The d.c. current  staircase is a result of the increasing number of the sidebands that are located within the bias voltage window $[\mu_L,\mu_R]$.
In general,  as the bias voltage window becomes bigger, there are more non-vanishing PAT sidebands contributing to the d.c. current in the WBL than in the SS-LSD model. 
In the latter, the memory effect quickly suppresses the higher $|m|$ PAT sidebands, leading to a much smaller d.c. current.
The insert in the upper corner illustrates how the SS-LSD ($b=1\ \hbar\omega$) staircase feature (solid curve) 
gets smoothed out in the WBL (dashed curve). 

\begin{figure}
\begin{center}
  \includegraphics[width=12.0 cm]{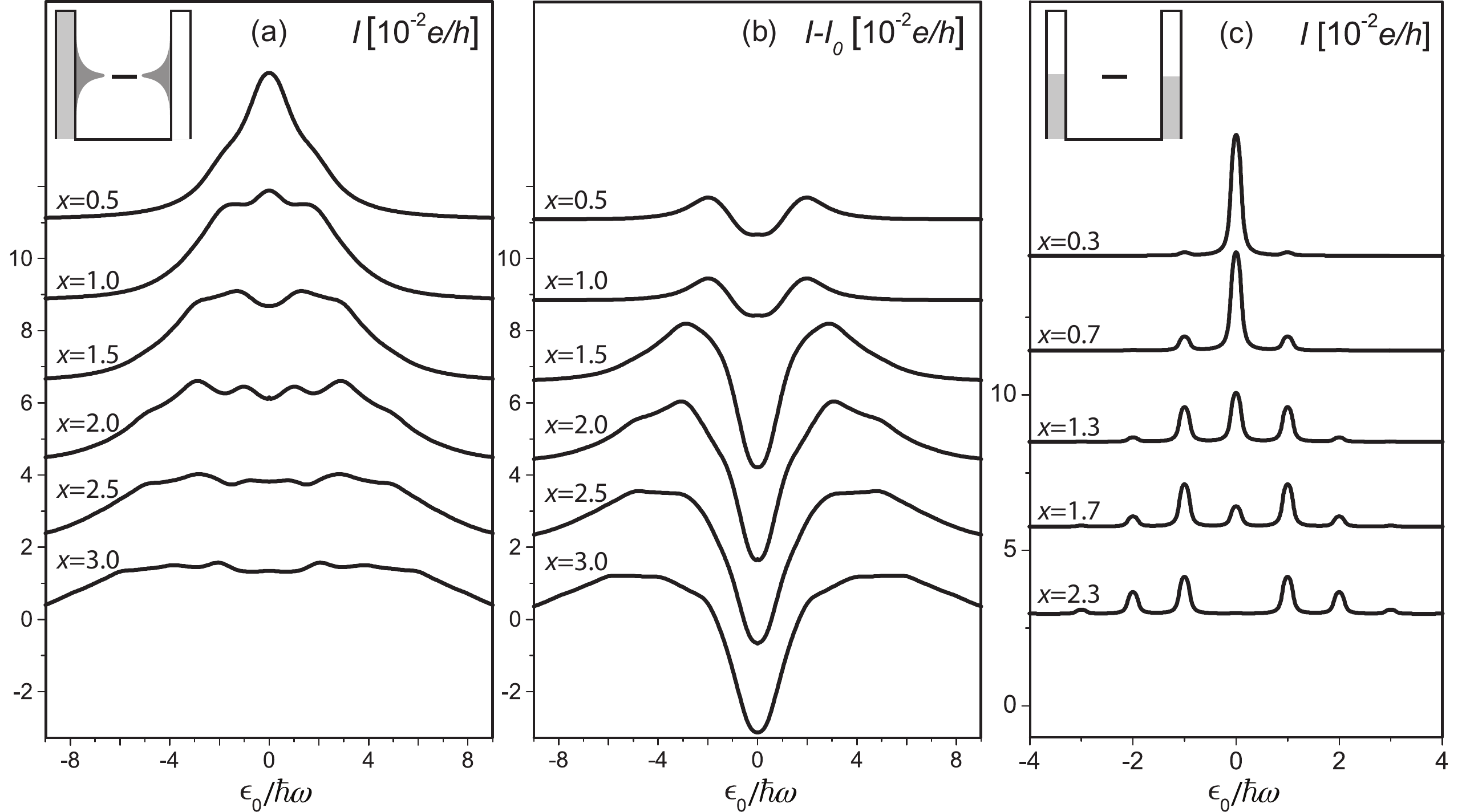}\\
  \caption{The d.c. currents $\bar{I}$ and the field-induced current $\bar{I}-\bar{I}_0$ are plotted as a function of the level energy ($\epsilon_{0}$) for a large bias voltage ($\mu_{L}=-\mu_{R}=10~[\hbar\omega]$) in the SS-LSD model ($\sigma=0$ and $b=1\ \hbar\omega$), shown in (a) and (b). The insets shows the corresponding energy diagrams. We also plot the d.c. current in the WBL for the zero bias voltage case ($\mu_{L}=-\mu_{R}=0.1~\hbar\omega$) in the panel (c). }
  \label{IE0X3}
\end{center}
\end{figure}

Figure~\ref{IE0X3} shows the memory effect in the SS-LSD model on the electron tunneling blockade phenomenon in the large bias voltage limit.  
Specifically, the d.c. current $\bar{I}$ calculated using a bias voltage window $[\mu_L=10\ \hbar\omega, \mu_R=-10\ \hbar\omega]$ is shown as a function of the reference gate voltage $\epsilon_0$ for different amplitudes $A$ of the external field (with $\sigma=0$ and $b=1\ \hbar\omega$ fixed in the calculations).
Fig.~\ref{IE0X3}(a) shows that 
(1) the d.c. current feature (as a function of $\epsilon_0$) strongly depends on the value of $x(=A/\hbar\omega)$ and 
(2) the d.c. current drops to zero as soon as the reference gate voltage $\epsilon_0$ moves outside the SS-LSD width $b$. The latter finding is responsible for the d.c. current blockade caused by the memory effect in the SS-LSD model. 
The amplitude dependent behavior seen in Fig.~\ref{IE0X3}(b) is complete absent in the WBL where the d.c. current at very large bias voltage can be computed explicitly as
\begin{equation*}
\bar{I}_{WBL}=\frac{e}{h}\frac{\pi\Gamma}{2},
\end{equation*}
which depends only on the coupling strength $\Gamma$ (here  $\bar{I}_{WBL}=5.772\times10^{-2}~[e/h]$ for $\Gamma=0.1\hbar\omega$).
By taking the memory effect into account, the tunneling current is effectively blocked between the SQD and the electrode 
because of the limited accessible energy bandwidth in the SS-LSD model.
The tunneling coupling between the SQD and the electrodes is strongest when the energy level $\epsilon_0$ coincides with $\sigma$, see the inset in Fig.~\ref{IE0X3}(a).  
In addition, Fig.~\ref{IE0X3}(b) demonstrates that there exist significant oscillatory Coulomb blockade features in the field-induced current $\bar{I}-\bar{I}_{0}$ where $\bar{I}_{0}$ is the d.c. current in the absence of external fields.
The large bias voltage limit considered in Fig.~\ref{IE0X3}(a) and~\ref{IE0X3}(b) clearly reveals a very different mechanism (due to the memory effect) for the d.c. current Coulomb blockade phenomena from the mechanism responsible for the small bias situation in the WBL, in which the electron can tunnel through the SQD only when one of the sideband resides inside the small bias voltage window, as depicted in Fig.~\ref{IE0X3}(c).
The interplay of the band width $b$ and the bias voltage window $[\mu_{L},\mu_{R}]$ may result in the enhancement or suppression of the d.c. current by manipulating  the gate voltage $\epsilon_0$.

\section{Conclusions}

In summary, we have developed a generalized Floquet approach, including the memory effect, for the treatment of electron transport process on a periodically-driven single quantum dot system with multiple noninteracting levels.
Of particular interest, we have considered the symmetric single-function Lorentzian spectral density (SS-LSD) model for the electrodes and derived analytical expressions for the transmission coefficient and the d.c. current under the multi-photon resonance condition by the generalized Van Vleck (GVV) nearly-degenerate perturbation theory. 
The Tien-Gordon formula has been extended to include the memory effect and the multi-photon resonance processes, in particular, resulting in an effective multi-level model of a single quantum dot (one single level with multiple sidebands). 
The memory effect on the transmission coefficient and the d.c. current has been analyzed at the nearly-degenerate resonance conditions and for different external field amplitudes.
Numerical simulations of the transmission coefficients have shown that some multi-photon PAT sidebands can be suppressed by the memory effect.
We have also shown that the memory effect on the staircase feature of the d.c. current is closely related to the sideband width narrowing.
Furthermore, It has been observed that the electron tunneling may be blocked  for large bias voltage case if the gate voltage is moved outside the band width of the electrode spectral density function.  
The field-induced current oscillation may be produced and manipulated by applying periodic external fields, thus, enabling the enhancement or suppression of the d.c. current at certain gate voltages.

\section*{Acknowledgments}
This work was partially supported by National Science Council of Taiwan (No. 97-2112-M-002-003-MY3) and National Taiwan University (No. 98R0045 and 99R80870). TSH was partially supported by the U.S. Department of Energy.  SIC was partially supported by U.S. National Science Foundation and Department of Energy.

\appendix

\section{Derivations of Eq.~(\ref{GVVHn}) and the corresponding single-particle propagator}

The time-dependent equation, Eq.~(\ref{rGSE'}), can be written in terms of the Floquet quasi-states $|\alpha n\rangle=|\alpha\rangle\times e^{i n\omega t}$ in extended Hilbert space, leading to the Floquet Hamiltonian
\begin{equation}\label{HF}
\langle\alpha m|\mathbf{H}_{F}|\beta n\rangle
=\langle\alpha|\mathcal{H}^{(m-n)}|\beta\rangle-m\hbar\omega\delta_{\alpha\beta}\delta_{mn}\nonumber
\equiv\mathbf{H}_{0}+\xi\mathbf{V}
\end{equation}
where
\begin{equation}\label{H0}
\fl\mathbf{H}_{0}=
\left(
  \begin{array}{ccccccc}
    \ddots &  &  &  &  &  &  \\
     & \mathcal{H}_0+2\hbar\omega\mathcal{I} & 0 & 0 & 0& 0 &  \\
     & 0& \mathcal{H}_0+\hbar\omega\mathcal{I} &0 & 0 & 0&  \\
     & 0 & 0 & \mathcal{H}_0 & 0& 0 &  \\
     & 0 & 0 & 0& \mathcal{H}_0-\hbar\omega\mathcal{I} & 0 &  \\
     & 0 & 0 & 0 & 0 & \mathcal{H}_0-2\hbar\omega\mathcal{I} &  \\
     &  &  &  &  &  & \ddots \\
  \end{array}
\right),
\end{equation}
and
\begin{equation}\label{V}
\mathbf{V}=
\left(
  \begin{array}{ccccccc}
    \ddots &  &  &  &  &  &  \\
     & \mathcal{V}^{(0)} & \mathcal{V}^{(1)} & \mathcal{V}^{(2)} & \mathcal{V}^{(3)} & \mathcal{V}^{(4)} &  \\
     & \mathcal{V}^{(-1)} & \mathcal{V}^{(0)} & \mathcal{V}^{(1)} & \mathcal{V}^{(2)} & \mathcal{V}^{(3)} &  \\
     & \mathcal{V}^{(-2)} & \mathcal{V}^{(-1)} & \mathcal{V}^{(0)} & \mathcal{V}^{(1)} & \mathcal{V}^{(2)} &  \\
     & \mathcal{V}^{(-3)} & \mathcal{V}^{(-2)} & \mathcal{V}^{(-1)} & \mathcal{V}^{(0)} & \mathcal{V}^{(1)} &  \\
     & \mathcal{V}^{(-4)} & \mathcal{V}^{(-3)} & \mathcal{V}^{(-2)} & \mathcal{V}^{(-1)} & \mathcal{V}^{(0)} &  \\
     &  &  &  &  &  & \ddots \\
  \end{array}
\right),
\end{equation}
where $\mathcal{I}$ being a $2\times2$ identity submatrix and $\mathbf{H}_0$ and $\mathbf{V}$ are, respectively, composed of $2\times2$ sub-blocks shown in Eq.~(\ref{H0n}) and Eq.~(\ref{Vn}).

By invoking the nearly degenerate GVV perturbation theory, an effective $n$-photon resonant $2\times2$ Hamiltonian $\mathcal{H}_{GVV}$ and its eigenstates can be expressed as 
\begin{equation}
\mathcal{H}_{GVV}=\sum_{m=0}^{\infty}\xi^{m}\mathcal{H}^{(m)},
\end{equation}
\begin{equation}
|\varphi_{\pm}\rangle=\sum_{m=0}^{\infty}\xi^{m}|\varphi_{\pm}^{(m)}\rangle
\end{equation}
In terms of extended Hilbert space, the zeroth order eigenstates are
\begin{equation}
|\varphi_{+}^{(0)}\rangle=|1,0\rangle,\  |\varphi_{-}^{(0)}\rangle=|2,0\rangle.
\end{equation}
The first-order perturbation terms are, respectively,
\begin{equation}
\mathcal{H}^{(1)}
=\left(
\begin{array}{cc}
  0 & i J_n(x) \\
  -i J_n(x) & 0 \\
\end{array}
\right),
\end{equation}
and, 
\begin{equation}
|\varphi^{(1)}_{+}\rangle
=\sum_{m\neq0}\frac{-i J_{n+m}(x)}{i b-m\hbar\omega}|2,m\rangle,\ 
|\varphi^{(1)}_{-}\rangle
=\sum_{m\neq0}\frac{i J_{n+m}(x)}{-i b-m\hbar\omega}|1,m\rangle.
\end{equation}
The second order perturbation terms for $\mathcal{H}_{GVV}$ can be written as
\begin{equation}
\mathcal{H}^{(2)}
=\left(
  \begin{array}{cc}
    -\delta_n & 0 \\
    0 &  \delta_n \\
  \end{array}
\right),
\end{equation}
where
\begin{equation}\label{delta_app}
\eqalign{
\delta_{n}&=\sum_{m\neq0}\frac{J_{n+m}^2(x)}{-i b+m\hbar\omega}\\
&=\left\{\begin{array}{lll}
i\sum\limits_{m\neq0}\frac{b}{(m\hbar\omega)^{2}+b^{2}}J^2_m(x)&\mbox{if}&n=0\\
\sum\limits_{m\neq0}\frac{m\hbar\omega}{(m\hbar\omega)^{2}+b^{2}}J^2_{n+m}(x)
+i\sum\limits_{m\neq0}\frac{b}{(m\hbar\omega)^{2}+b^{2}}J^2_{n+m}(x)&\mbox{if}&n\neq 0\\
\end{array}\right.
}
\end{equation}
Hence, the GVV effective Hamiltonian, up to the second-order, can then take on the form
\begin{equation}\label{GVVHnA}
\mathcal{H}_{GVV}\approx
\left(
  \begin{array}{cc}
    \epsilon_{0}-\xi^{2}\delta_{n} & i\xi J_n(x) \\
    -i\xi J_n(x) & \epsilon_{0}-i b+\xi^{2}\delta_{n}  \\
  \end{array}
\right),
\end{equation}
compared to the corresponding effective Hamiltonian in the rotating wave approximation (RWA),
\begin{equation}
\mathcal{H}_{RWA}\approx
\left(
  \begin{array}{cc}
    \epsilon_{0} & i\xi J_n(x) \\
    -i\xi J_n(x) & \epsilon_{0}-i b \\
  \end{array}
\right).
\end{equation}

In the weak coupling limit and with a finite band width, the eigenvalues of the GVV effective Hamiltonian $\mathcal{H}_{GVV}$ can be approximated as
\begin{equation}
\left\{\begin{array}{l}
\lambda_+\approx\epsilon_0-\Delta_n-i\Xi_n \\
\lambda_-\approx\sigma-n\hbar\omega-i b+\Delta_n+i\Xi_n.
\end{array}\right.
\end{equation}
where
\begin{equation}
\Delta_{n}\equiv\left\{\begin{array}{lll}
0&\mbox{if}& n=0\\
\frac{\Gamma}{2}\sum\limits_{m\neq 0}\frac{b m\hbar\omega}{(m\hbar\omega)^{2}+b^{2}}J^2_{n+m}(x)&\mbox{if}& n\geq 1\\
\end{array}\right.
\end{equation}
and
\begin{equation}
\Xi_{n}\equiv\frac{\Gamma}{2}\sum_{m=-\infty}^{\infty}\frac{b^2}{(m\hbar\omega)^{2}+b^{2}}J^2_{n+m}(x),
\end{equation}
Hence, we can derive the fundamental solution by $\mathcal{U}(t,t')\approx\mathcal{R}(t)\mathcal{U}_{GVV}(t,t')\mathcal{R}^{\dagger}(t')$ and the single-particle propagator $U(t,t')$, cf. Eqs. (\ref{rotationRn}) and (\ref{UU}),
\begin{equation}
U(t,t')\approx e^{-i\Omega(t)}e^{-i\lambda_+(t-t')/\hbar}e^{i\Omega(t')}.
\end{equation}
In conjunction of $e^{i x\sin\omega t}=\sum J_m(x)e^{i m\omega t}$, we derive the Green's function, cf. Eq.~(\ref{Gte}),
\begin{equation}
G(t,\epsilon)\approx\sum_{k=-\infty}^{\infty}\sum_{m=-\infty}^{\infty}
\frac{J_m(x)J_{m+k}(x)}{\epsilon-\lambda_+-m\hbar\omega}\times e^{-i k\omega t},
\end{equation}
and its $k$-th order Fourier coefficients, cf. Eq.~(\ref{Gke}), are 
\begin{equation}\label{Gke3}
G^{(k)}(\epsilon)\approx\sum_{m=-\infty}^{\infty}
\frac{J_m(x)J_{m+k}(x)}{\epsilon-\epsilon_{0}-m\hbar\omega+\Delta_n+i\Xi_n}.
\end{equation}

\section*{References}

\bibliographystyle{IOP_bst}
\bibliography{ArXiv_draft}

\end{document}